\documentclass[aps,twocolumn,superscriptaddress,nofootinbib]{revtex4-1} 
\usepackage{dcolumn}    
\usepackage{pstricks}
\usepackage{color}
\usepackage{graphicx}
\usepackage{amsmath}
\usepackage{amssymb}
\usepackage[colorlinks=true,allcolors=darkred, pdfborder={0 0 0}]{hyperref}

\usepackage{booktabs}
\renewcommand{\toprule}{\specialrule{1.5pt}{0em}{0em}}
\renewcommand{\bottomrule}{\specialrule{1.5pt}{0em}{0em}}

\def\five{\mathbf{5}}
\def\fiveb{\mathbf{\overline{5}}}
\def\fives{\mathbf{5^\star}}
\def\ten{\mathbf{10}}
\def\fortyf{\mathbf{45}}
\def\twentyf{\mathbf{24}}

\def\nn{\nonumber}

\def\vev#1{\left\langle #1\right\rangle}

\newcommand{\fig}[1]{Figure~(\ref{fig:#1})}
\newcommand{\Fig}[1]{Figure~(\ref{fig:#1})}

\newcommand{\Tab}[1]{Table~(\ref{tab:#1})}
\newcommand{\Sec}[1]{Sec.~(\ref{sec:#1})}

\newcommand{\eq}[1]{eq.~(\ref{eq:#1})}

\definecolor{darkred}{rgb}{0.6,0,0}

\begin{document}

\title{Axion inflation, proton decay, and leptogenesis in $SU(5)\times U(1)_{PQ}$}

\author{Sofiane M. Boucenna}
\email{boucenna@kth.se}
\affiliation{Department of Physics, School of Engineering Sciences, KTH Royal 
Institute of Technology, AlbaNova University Center, Roslagstullsbacken 21, 
SE–10691 Stockholm, Sweden}
\affiliation{The Oskar Klein Centre for Cosmoparticle Physics,
AlbaNova University Center, Roslagstullsbacken 21, SE–10691 Stockholm, Sweden}
\author{Qaisar Shafi}
\email{shafi@bartol.udel.edu}
\affiliation{Bartol Research Institute, University of Delaware, Newark, DE 19716 
USA.}

\begin{abstract}
We implement inflation in a non-supersymmetric
$SU(5)$ model based on a non-minimal coupling of the axion field
to gravity. The isocurvature fluctuations are adequately suppressed,
axions comprise the dark matter, proton lifetime estimates are of order $8 \times 10^{34}-3 \times 10^{35}$ yr, and the observed baryon asymmetry
arises via non-thermal leptogenesis. The presence of low scale colored scalars
ensures unification of the Standard Model gauge couplings and also helps in stabilizing the electroweak vacuum.
\end{abstract}
 \date{\today}

\maketitle

\section{Introduction}
\label{sec:into}

Successful inflation models based on $SU(5)$ grand unified theory (GUT) and employing a Coleman-Weinberg
potential with minimal coupling to gravity were constructed some time
ago \cite { ShafiVilenkinSU5} and, for values of the scalar spectral index 
$n_{s}\sim 0.96-0.97$,
the tensor to scalar ratio $r$ is $\gtrsim 10^{-2}$ \cite{Rehman:2008qs}. In these models
the inflaton field, an $SU(5)$ gauge singlet, evolves from the origin
to its non-zero vev, reaching trans-Planckian values near its minimum
during the last 60 or so e-foldings.
Identifying the $SU(5)$ gauge singlet inflaton with the axion \cite{Wilczek:1977pj,Weinberg:1977ma} field
is very attractive and was first done in \cite{Pi:1984pv}. However, it turns out to be not very compelling for this model because
of the excessively large value $f_{a}\sim10^{19}$ GeV imposed on
the axion decay constant by inflation \cite{Rehman:2008qs}. Typically, an axion decay constant
$f_{a}\sim10^{11-12}$ GeV is the desired value for axion dark matter.\\

It has been known for a while  that primordial inflation driven
by a scalar quartic potential and based on non-minimal coupling to
gravity is fully consistent with the {\sc Planck}  observations
\cite{Okada:2014lxa} for plausible values, say $\xi \lesssim 10$, of the 
dimensionless
non-minimal coupling parameter $\xi$. The inflaton field in this
case rolls down from trans-Planckian values to its final minimum which
can be sub-Planckian as desired. The scalar quartic coupling $\lambda$
during inflation in this case turns out to be $\lesssim10^{-8}$.
This means that in order to protect $\lambda$ from unacceptably large
radiative corrections, in non-supersymmetric GUTs the inflaton should
again be identified with a gauge singlet scalar field, as was done
in the $SU(5)$ model mentioned above.\\

An axion model needed to resolve the strong CP problem provides a
compelling candidate to implement successful inflation in GUTs using
non-minimal coupling to gravity. 
For the sake 
of simplicity, we will employ a Higgs rather
than the Coleman-Weinberg potential. 
The inflaton (radial component of the axion field) in this case
rolls down from trans-Planckian values during inflation to its final
value $f_a\sim 10^{11-12}$ GeV, thus yielding a viable scenario
with axion dark matter.\\

With trans-Planckian field values during inflation the isocurvature
fluctuations are adequately suppressed and observable gravity waves
corresponding to $r\sim \mathrm{few} \times 10^{-3}$ are predicted. The reheating process 
proceeds via the decay of the
inflaton into right-handed (RH) neutrinos. In turn, the latter yield the
observed baryon asymmetry via non-thermal leptogenesis.\\

The realistic GUT model we propose successfully addresses several problems of the SM at once, namely the existence and nature of dark matter, the strong CP problem, baryogenesis, the stability of the electroweak vacuum, the origin of the inflationary phase, and the physics behind neutrino masses. All of these issues have been previously studied in the literature in the aim of providing unified schemes which tackle several of them simultaneously, see e.g.,  Refs \cite{Langacker:1986rj,Shin:1987xc,Davoudiasl:2004be,Shaposhnikov:2006xi,Boucenna:2014uma,Bertolini:2014aia,Clarke:2015bea,Ahn:2015pia,Salvio:2015cja,Ballesteros:2016euj}. Here we show that a simple non-supersymmetric GUT model provides an elegant framework to solve all these problems, in addition to providing matter and gauge coupling unification.\\

This letter is organized as follows. We describe our model and outline its most salient features in \Sec{model}.
We then analyse the constraints on  gauge coupling unification (GCU) and proton decay in \Sec{masses}. These end up predicting the presence of a colored octet scalar not far from the TeV scale. Such a field also plays a critical
role in stabilizing the electroweak vacuum which we analyse in \Sec{vac}.
After which we outline the main features of inflation based on a quartic potential with non-minimal coupling
of the inflaton field to gravity in \Sec{inflation}, and  derive  the predictions for the spectral index $n_{s}$, the scalar-to-tensor ratio $r$, and the running of the spectral index $\alpha$. We also present
constraints on the magnitude of Yukawa couplings involving RH
neutrinos and estimate the reheat temperature taking into account
the requirement of axion dark matter. Next, in  in \Sec{lepto} we describe how leptogenesis
is implemented in this framework. Finally, in \Sec{monopoles} we show how the inflaton coupling to the adjoint $\mathbf{24}$-plet insures $SU(5)$ breaking during inflation so that the superheavy monopoles are
inflated away.

\begin{table}
\centering
\begin{tabular}{c|c|c|c||c|c|c|c|c}
\toprule
& $T_{L}$ & $F_{L}$ & $\nu_{L}^{c}$ & $H_{1}$ & $H_{2}$ & $\sigma$ & $\Phi$ & 
$\chi$\tabularnewline
\hline
$SU(5)$ & $\ten$ & $\fiveb$ & $\mathbf{1}$ & $\five$ & $\fiveb$ & $\mathbf{1}$ 
&$\twentyf$ &$\fortyf^\star$\tabularnewline
\hline
$U(1)_{PQ}$ & $q/2$ & $q/2$ & $q$/2 & $-q$ & $-q$ & 
$-q$ & 0 &  $-q$\tabularnewline
\hline
\hline
$U(1)_{X}$ & $1/5$ & $-3/5$ & $-1$ & $-2/5$ & $2/5$ & 
$2$ & 0 &  $2/5$\tabularnewline
\bottomrule
\end{tabular}
\caption{Summary of the quantum numbers of the different fields of the 
model. $q$ is an arbitrary number. $U(1)_X$ is an accidental symmetry of the model when the mixed term $(\sigma^\star)^2 H_1 H_2$ is absent, i.e., in the limit $\lambda_\nu\to 0$.}\label{tab:content}
\end{table}

\section{The model}
\label{sec:model}

The model consists of a simple extension of the original $SU(5)$ model~\cite{Georgi:1974sy}. The SM 
fermion fields are in the usual $\ten$ ($T_L$) and $\fiveb$ ($F_L$), and we add 
the singlet RH neutrinos $\nu_L^c$. The scalar sector of the model 
involves   $\five$ ($H_1$), $\fives$ ($H_2$),  $\twentyf$ ($\Phi$), and 
lastly  $\fortyf^\star$ ($\chi$, with $\chi^{ij}_k=-\chi^{ji}_k$ for $i,j,k=1-5$). In addition, we define a global $U(1)_{PQ}$ 
symmetry to implement the Peccei-Quinn (PQ) mechanism~\cite{Peccei:1977ur} solving the strong CP problem and
providing an invisible axion via the DFSZ mechanism~\cite{Zhitnitsky:1980tq,Dine:1981rt};\footnote{For an early reference on an explicit $SU(5)$ construction solving the strong CP problem we refer to Ref.~\cite{Wise:1981ry}.} see \Tab{content} for the $U(1)_{PQ}$  charges of the different fields.\\

The relevant terms in the scalar potential of the model read:
\begin{eqnarray}
V& \supset & -\tfrac{1}{2}M_{GUT}^{2}\mathrm{{Tr}}(\Phi^{2}) -  M_\sigma^2 \sigma\sigma^\star +  \lambda_\sigma (\sigma^\star \sigma)^2 \nn\\
&& + \sigma\sigma^\star\sum_{\phi} \kappa_\phi \phi^\star\phi+ \lambda_\nu [ (\sigma^\star)^2 H_1 H_2 +\mathrm{h.c.}]\,.
\label{eq:V}
\end{eqnarray}
Here $\phi=(H_1,H_2, \Phi, \chi)$ and the dot product $ \phi^\star.\phi$ 
represents the $SU(5)\times U(1)_{PQ}$ invariant contractions. And the Yukawa part of the Lagrangian is given by the following 
terms (family, gauge, and Lorentz indices  are suppressed):
\begin{eqnarray}
\mathcal{L}_{Yuk} &=& T_L.\mathbf{Y_{10}}.T_L.H_1 +  T_L.\mathbf{Y_5}.F_L.H_2 
+ T_L.\mathbf{Y_{45}}.F_L.\chi\,\nn\\
&& F_L.\mathbf{Y_\nu}.\nu_L^c.H_1 + \tfrac{1}{2} \mathbf{Y_N} 
\nu_L^c.\nu_L^c.\sigma + \mathrm{h.c.}\,,
\label{eq:yuks}
\end{eqnarray}
where $\mathbf{Y}_i$ are dimensionless $3\times 3$ matrices. 
The PQ symmetry is tightly related to lepton number in this scheme. We can 
readily see that if $\lambda_\nu=0$, the Lagrangian is invariant under  
$U(1)_{B-L}$ symmetry after the breaking of $SU(5)$ and $U(1)_{PQ}$. Indeed, an accidental
$U(1)_X$ symmetry (shown in \Tab{content}) combines with the usual hypercharge to leave an unbroken $U(1)$ defined by
$X+\tfrac{4}{5}Y \equiv B-L$. However, this symmetry is explicitely broken in the scalar sector due to the presence of the $\lambda_\nu$ term which is crucial for generating the axion in the DFSZ model and cannot be set to zero.  Additionally, $\lambda_\nu \neq 0$ allows  us to get rid of the majoron  goldstone boson \cite{Chikashige:1980ui,Gelmini:1980re} since lepton number is broken explicitely~\cite{Langacker:1986rj}.\\

The  representations involved in the model play crucial roles in different 
phenomenological sectors. Namely,
\begin{itemize}
\item $H_{1,2}$ and $\chi$ account for the SM charged fermion masses and 
mixings in a renormalizable way. The two-Higgs-Doublet model (2-HDM) consisting of $H_{1,2}$ ensures the 
implementation of the DFSZ mechanism. The multiplet $\chi$  is also crucial for 
obtaining accurate gauge coupling unification;
\item $\vev{\Phi}$ breaks $SU(5)$ to the SM;
\item the phase of $\sigma$ is the axion which solves the strong CP 
problem and accounts for the DM of the universe, and the radial part drives inflation;
\item $\vev{\sigma}$ provides large Majorana masses for $\nu_L^c$ via the see-saw mechanism. After inflation the latter helps 
generate the observed baryon asymmetry via non-thermal leptogenesis.
\end{itemize}

In the next sections we will investigate in details all these aspects of the 
model.

\section{Fermions masses, gauge coupling unification, and proton decay}
\label{sec:masses}

From \eq{yuks}, we obtain the following mass relations:
\begin{eqnarray}
M_e &=& \mathbf{Y_5}^T \vev{H_2} -6\mathbf{ Y_{45}}^T \vev{\chi}\\
M_d &=&\mathbf{Y_5} \vev{H_2} + 2 \mathbf{Y_{45}} \vev{\chi}\\
M_u &=& 4 (\mathbf{Y_{10}} + \mathbf{Y_{10}}^T) \vev{H_1} \\
M_\nu &\simeq& \mathbf{Y_\nu}^T.\mathbf{Y_N}^{-1}.\mathbf{Y_\nu} 
\tfrac{\vev{H_1}^2}{\vev{\sigma}}\,.
\end{eqnarray}

In the last equation we have assumed the see-saw scaling $\vev{\sigma} \gg 
\vev{H_1}$ for natural couplings. We define $\vev{\chi}\equiv \vev{\chi}^{15}_1=\vev{\chi}^{25}_2=\vev{\chi}^{35}_3=-3\vev{\chi}^{45}_4$. It is clear from these expressions that there  
is enough parameter freedom to fit the fermion masses and cure the wrong 
predictions of minimal $SU(5)$ \cite{Georgi:1979df,Kalyniak:1982pt,Eckert:1983bn}.\\

Next, we turn to the issue of gauge coupling unification 
(GCU). In the minimal $SU(5)$ model, the gauge couplings do not properly unify at high energy. However, the 
$\chi$ and $\Phi$ multiplets contain representations which can alter the renormalization group (RG) evolution in a favourable way~\cite{Dorsner:2006dj}.
In particular, we will use $R_8  \equiv 
(\mathbf{8},\mathbf{2},1/2) \in \chi$ and $R_3 \equiv 
(\mathbf{3},\mathbf{3},-1/3) \in \chi$ to obtain precise GCU. Note that in the absence of the PQ symmetry $R_3$ can mediate proton decay leading to a lower limit on its mass that was estimated to be around $10^{10}$ GeV ~\cite{Dorsner:2006dj}. However, in our scenario $R_3$ cannot induce nucleon decay  due to the absence of the couplings $T_L.\chi.H_2$ or $T_L.\chi.H_1^\star$ and thus it can be very light. This significantly enlarges the parameter space consistent with GCU.\\

We solve the system of RG equations in order to obtain successful GCU. The 
equations depend on 3 parameters: $M_{R_3}$,  $M_{R_8}$, and  $M_{GUT}$ (for 
simplicity, we ignore threshold effects). We require that $M_{GUT}$ is large 
enough so that gauge-mediated proton decay does not  rule out the model, i.e., 
\begin{eqnarray}
\tau_p \sim \alpha_{GUT}^{-2} \frac{M_{GUT}^4}{m_p} \gtrsim  10^{34} \, \mathrm{yr}\,,
\end{eqnarray}
where $m_p$ is the proton mass and the lower limit is the current experimental bound on $\tau_p (p\to e^+ \pi^0)$\cite{PDG2016}.
After combining these constraints, we find:
\begin{eqnarray}
M_{GUT} &\approx& \left(\frac{M_{R_8}}{\mathrm{TeV}} \right)^{-0.126} \times  
10^{16}\, \mathrm{GeV} \,,\label{eq:mgut}\\ 
M_{R_3} &\approx& \left(\frac{M_{R_8}}{\mathrm{TeV}}\right)^{0.05} \times 6.1 \times
10^7 \, \mathrm{GeV} \,,
\end{eqnarray}
and
\begin{eqnarray}
M_{R_8} &\lesssim& 6 \times 10^5 \, \mathrm{GeV}\,.
\end{eqnarray}

\begin{figure}[t!]
\centering
 \includegraphics[clip,width=0.48\textwidth]{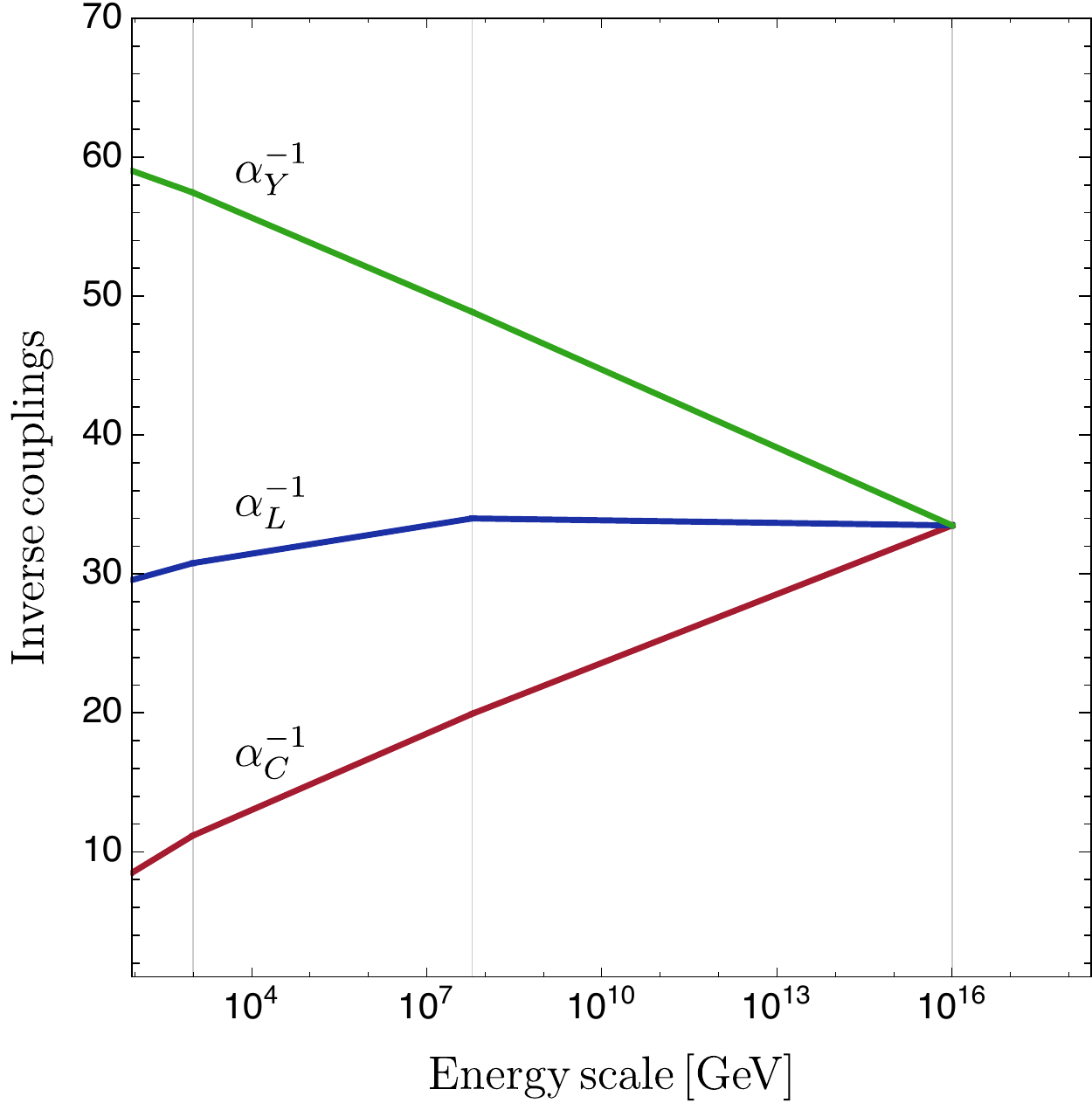}\caption{Evolution of the inverse coupling strengths with energy. Here $M_{R_8} = 1$ TeV, $M_{R_3}\approx 6.1\times 
10^7$ GeV, and $M_{GUT}\approx 10^{16}$ GeV. The proton lifetime in this case is $\tau_p \approx 2.4\times 10^{35}$ yr.}\label{fig:gcu}
\end{figure}

The maximum proton lifetime is achieved for the smallest possible $R_8$ mass. For 1 TeV mass, we obtain $\tau_p \approx 2.4\times 10^{35}$ yr, which is around the expected sensitivity of Hyper-Kamiokande experiment~\cite{Abe:2011ts}.
We display the gauge coupling unification in \Fig{gcu} for the case where $M_{R_8} = 1$ TeV.

\section{Vacuum stability}
\label{sec:vac}

\begin{figure}[t!]
\centering
 \includegraphics[clip,width=0.48\textwidth]{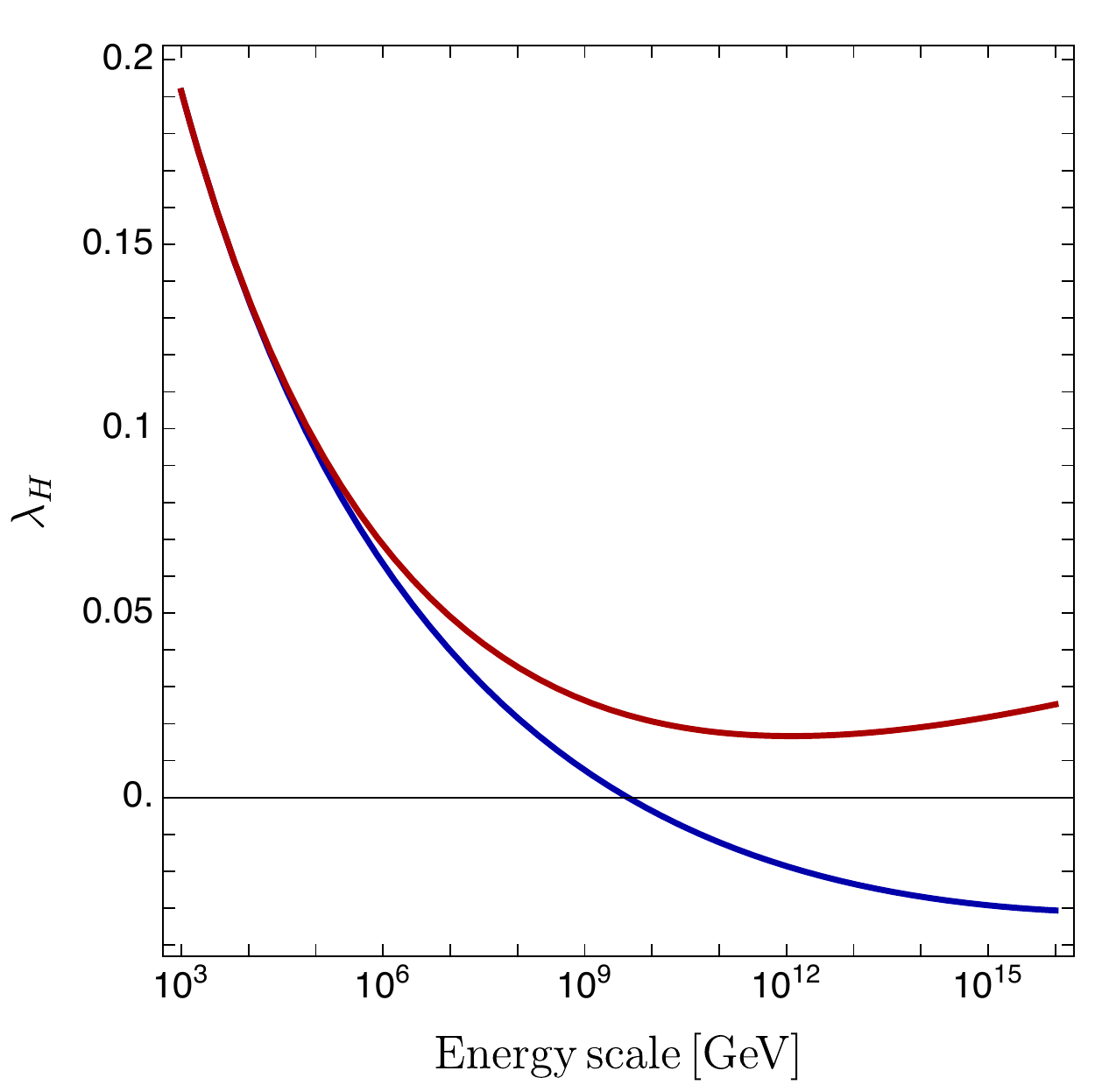}
 \caption{Vacuum stability of the Higgs potential. The red (upper) line shows the running of the Higgs quartic coupling in the model compatible with \fig{gcu}. We display in blue (lower line) its running in the SM for comparison.}
 \label{fig:vac}
\end{figure}

It is well known that that the vacuum instability problem associated with the quartic coupling of the Higgs  (see for instance~\cite{Buttazzo:2013uya}) can be overcome with new physics around the TeV scale. The GCU analysis in the previous section revealed that the model predicts scalar representations which have to be  lighter than $\sim 10^{10}$ GeV, more or less the scale at which the quartic coupling becomes negative. This is remarkable as such scalars will contribute to the running of the quartic coupling and could positively tilt it before it becomes negative. In this section we will analyse the effect of these scalars on the stability of the vacuum.
For renormalization energy $\mu < M_{R_8}$, we use the SM RG equations at two-loop level to calculate the evolution of the Higgs quartic coupling~\cite{Machacek:1983tz,Machacek:1984zw,Ford:1992pn,Arason:1991ic,Barger:1992ac,Luo:2002ey}. We include the effects from the new particles $R_3$ and $R_8$ at one-loop level. These modify the first order coeffcients  $b_i$  of the SM, $\tfrac{d g_i}{d ln\mu} = \tfrac{b_i}{16 \pi^2} g_i^3$,  $b_i (\mu)= (\tfrac{41}{10},-\tfrac{19}{6},-7) + \Theta(\mu-M_{R_8}) (2,\tfrac{4}{3},\tfrac{4}{5})+\Theta(\mu-M_{R_3}) (\tfrac{1}{2},2,\frac{1}{5})$.
In solving the RGEs, we use the boundary conditions at the top quark pole mass given in~\cite{Buttazzo:2013uya}. For the $SU(3)_c$ coupling constant and the top mass, we use $\alpha_s=0.1184$ and $M_t=173.34$ GeV~\cite{ATLAS:2014wva} respectively. The SM Higgs mass is fixed at $M_h=125.09$ GeV~\cite{Aad:2015zhl}. We find in particular that $R_8$, being very  light, induces a significant effect on the running of the gauge couplings.  The scalar $R_3$ on the other hand has a negligible effect on the running before the instability scale.
As  we can see in \fig{vac}, the quartic Higgs coupling is indeed prevented from becoming negative by including the $R_8$ field at 1 TeV and $R_3$ at $6.1\times 10^7$ GeV. Remarkably, the same fields which alllow us to implement GCU also stabilise
the effective potential of the SM at high energies.\\

Finally, we can use this analysis to constrain the mass of $R_8$. Indeed, the heavier it is the less important is its effect on the Higgs quartic coupling, and so we expect an upper bound not far from the TeV region. We find that
\begin{eqnarray}
M_{R_8} &\lesssim& 10^4 \, \mathrm{GeV}\,,
\end{eqnarray}
which is more constraining than the upper bound derived from GCU considerations only. Using \eq{mgut}, this upper bound translates as $\tau_p \gtrsim 7.8\times 10^{34}$ yr.

\section{Axion Inflation}
\label{sec:inflation}

We assume that inflation is driven by the radial part of the complex 
singlet $\sigma$, $\rho\equiv\sqrt{{2}}\mathcal{{R}}e(\sigma)$. Without loss of generality, we take
$\mathbf{Y_N }~=~\mathrm{diag(Y_{N_1},Y_{N_2},Y_{N_3})}$,  with $Y_{N_i}$ real and positive. The relevant terms of the Lagrangian of the model are:
\begin{eqnarray}
\mathcal{{L}}_{inf} & = & 
\left(\sum_{i=1}^3\frac{Y_{N_i}}{2\sqrt{2}}\,\rho\,{\nu_{iL}^{c 
}}\nu_{iL}^c+h.c.\right)-V_{inf}\,,\\
V_{inf} & = &  \tfrac{1}{4}\lambda_\sigma (\rho^2 - f_a^2)^2 +\tfrac{1}{2}\rho^2 \sum_{\phi} 
\kappa_\phi \phi^\star.\phi\,,
\label{eq:Linf}
\end{eqnarray}
where $f_a \equiv \sqrt{2}\vev{\sigma}$ and for simplicity, 
we only consider real couplings. We enforce $\kappa_{H_{1,2}}>0$ to ensure that 
inflation is driven by $\rho$. The  couplings of the inflaton with 
$\nu_{L}^c$ and the scalar fields induce quantum corrections that can have 
significant effects on the inflationary observables. These effects induce an 
additional contribution to the potential $V_{inf}$ denoted as $V^{(1)}$,
\begin{equation}
\label{eq:cw}
V^{(1)}=\frac{\beta}{16\pi^2}\,\rho^4 \mathrm{ln}\frac{\rho^2}{\mu^2}\,,
\end{equation}
where \cite{Okada:2013vxa}
\begin{equation}
\label{eq:beta}
\beta=20\lambda_\sigma^{2}+2\sum_\phi\kappa_\phi^{2}+2\lambda_\sigma\sum y_{N 
i}^{2}-\sum y_{N i}^{4}\,.\nn
\end{equation}
We require that these radiative corrections are not significant, i.e., 
$|\beta|\ll (16\pi^2)|\lambda_\sigma|$.  The most conservative limit we can set 
on the couplings is (defining max($Y_{N_i}) = Y_N$): 
\begin{equation}
y_{N} \lesssim 6\times 10^{-2}\,\left(\frac{\lambda_\sigma}{10^{-7}} 
\right)^{\tfrac{1}{4}}\,.
\label{eq:yN}
\end{equation}
For the rest of the paper we will suppose that $\kappa_\phi \ll y_N$ and impose \eq{yN}.

\subsection{$\rho^{4}$ inflation with non-minimal coupling to gravity}
We consider a scenario where $\rho$ has a non-minimal coupling to gravity. For 
simplicity, we  assume that all other scalars, including the SM Higgs, have 
quasi-minimal couplings. In the Jordan frame, the action of non-minimal 
$\rho^{4}$ inflation
is given by:

\begin{equation}
S_{J}^{{\rm tree}}=\int 
d^{4}x\sqrt{-g}\left[-\left(\frac{1}2(1+\xi\rho^{2})\right)\mathcal{R}+\frac{1}{
2}(\partial\rho)^{2}-\frac{\lambda_\sigma}{4}\rho^{4}\right]\,.
\end{equation}

In the Einstein frame one finds,
\begin{equation}
S_{E}=\int 
d^{4}x\sqrt{-g_{E}}\left[-\frac{1}{2}\mathcal{R}_{E}+\frac{1}{2}(\partial S)^{
2}-V_{E}(S(\rho))\right]\,,
\end{equation}
where the canonically normalized scalar field $S$ is written in
terms of the original scalar as:
\begin{equation}
\left(\frac{dS}{d\rho}\right)^{-2}=\frac{\left(1+\xi\rho^{2}\right)^{2}}{
1+(6\xi+1)\xi\rho^{2}}\,.
\end{equation}
The inflation potential now reads:
\begin{equation}
V_{E}(S(\rho))=\frac{\frac{1}{4}\lambda_\sigma(t)\rho^{4}}{\left(1+\xi\,\rho^
{2}\right)^{2}}\,,
\end{equation}
and the inflationary slow-roll parameters~\cite{Liddle:1992wi,Liddle:1993fq} in terms of  $\rho$ are expressed as: 
\begin{eqnarray*}
\epsilon(\rho) & = & \frac{1}{2}\left(\frac{V_{E}'}{V_{E}S'}\right)^{2},\\
\eta(\rho) & = & 
\frac{V_{E}''}{V_{E}(S')^{2}}-\frac{V_{E}'S''}{V_{E}(S')^{3}},\\
\frac{\zeta(\rho)}{\sqrt{2\epsilon(\rho)}} & = & 
\frac{V_{E}'''}{V_{E}(S')^{3}}-\frac{3 
V_{E}''S''}{V_{E}(S')^{4}}+\frac{3 
V_{E}'(S'')^{2}}{V_{E}(S')^{5}}-\frac{V_{E}'S'''}{V_{E}(S')^{4}}\,,
\end{eqnarray*}
where a prime denotes derivative with respect to $\rho$, and we use units where the reduced Planck 
mass, $M_{\mathrm{Pl}}\simeq 2.4\times 10^{18}$ GeV, is equal to unity unless otherwise stated. The number of e-folds is given by:
\begin{equation}
N=\frac{1}{\sqrt{2}}\int_{\rho_{{\rm 
e}}}^{\rho_{0}}\frac{d\rho}{\sqrt{\epsilon(\rho)}}\left(\frac{d S}{d\rho}
\right)\,.
\end{equation}

The inflationary predictions for  the scalar spectral index $n_s$, the tensor-to-scalar ratio $r$, and the running of the spectral index $\alpha=\tfrac{\mathrm{d}n_s}{\mathrm{d ln}k}$ are
obtained after fixing $N$ and $\xi$. The quartic coupling $\lambda_\sigma$
can be fixed using the amplitude of density perturbations at some
pivot scale \cite{Ade:2015lrj},
\begin{equation}
\Delta_{\mathcal{{R}}}^{2}=\left.\frac{V_{E}}{24\pi^{2}\epsilon}\right|_{k^{
\star}}=\left.2.196\times10^{-9}\right|_{0.05\mathrm{{Mpc}^{-1}}}\,.
\end{equation}
 
\begin{figure}[t]
\centering
\includegraphics[clip,width=0.45\textwidth]{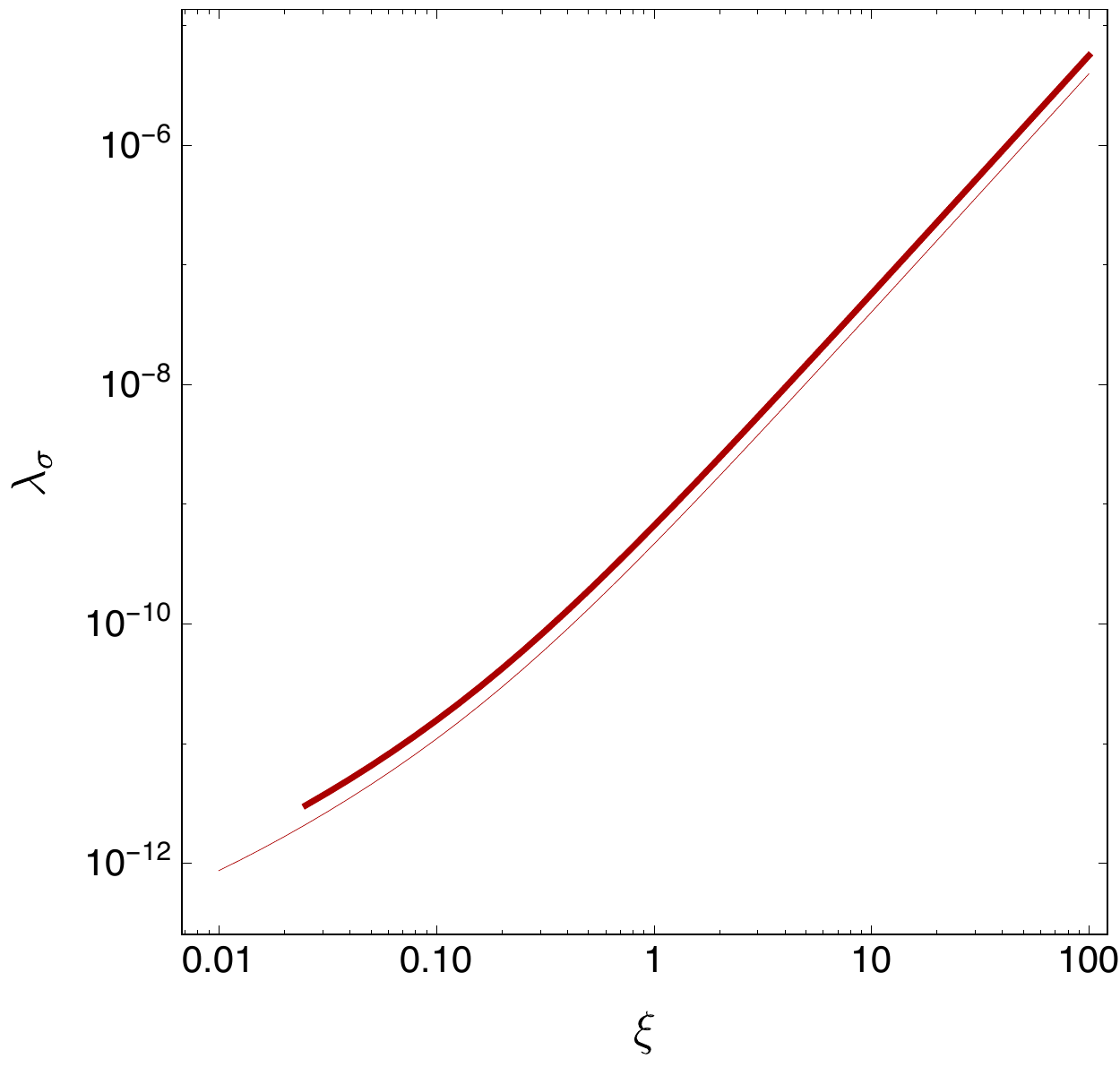}\caption{
Correlation between the inflaton's quartic coupling $\lambda_\sigma$ and the non-minimal 
coupling $\xi$ for $N=50$ e-foldings (thick line) and $N=60$ e-foldings (thin line). $n_s$ is within the 68\% confidence level of {\sc Planck}'s measurement. }\label{fig:lvsxi}
\end{figure}

In \Fig{lvsxi}  we show the predicted values of the quartic coupling as a function of the minimal coupling $\xi$ for $N=50$ and $N=60$ e-foldings. $n_s$ is constrained to be  within the 68\% confidence level of {\sc Planck}'s measurement \cite{Ade:2015lrj}.
 
For $\xi \gtrsim 0.1$, the predicted values of $n_s$, $r$, and $\alpha$ quickly converge toward:

\begin{center}
\begin{tabular}{c|c|c|c }
& $n_s$ & $r \times 10^3$ &$- \alpha\times 10^4$\\
\hline
$N=50$ & $0.962$ & $4$  & $7.5$\\
$N=60$ & $0.968$ & $3$ & $5.3$\\
\hline
\end{tabular}
\end{center}

This implies that the Hubble expansion rate at the end of 
inflation is:
\begin{equation}
H_{I}\simeq2\pi\times10^{13}\;\mathrm{{GeV}\,}\,.
\label{eq:Hinf}
\end{equation}

\subsection{Reheating}
\label{sec:trh}

As can be seen in \eq{yN}, the coupling of the inflaton with $\nu^c_L$ can be 
sizable, and this can be used to reheat the Universe  via the decays $\rho \to 2 
\nu^c_L$. This is the dominant process because our assumption that 
$\kappa_{\phi} \ll 1$ makes  reheating via the scalars inefficient. In order 
to do so, the mass of at least one of RH neutrinos $M_{N}= y_{N}\,f_{a}/\sqrt{2}$  must be 
smaller than half the inflaton's mass $m_{\rho}=B\,f_{a}$, with $B\simeq 
\sqrt{2\lambda_\sigma}$. This translates as:
\begin{equation}
y_{N}<\sqrt{2\lambda_\sigma}\label{eq:yN2}\,.
\end{equation}

Note that this condition is more stringent than the one obtained in \eq{yN}.
Assuming an instantaneous conversion of the inflaton\textquoteright s
energy density into radiation, at the time when $H(t)\approx\Gamma_{\rho}$ (decay rate of $\rho$),
we can define the reheating temperature as:
\begin{equation}
T_{RH}=\left(\frac{45}{4\pi^{3}g_{\star}}\right)^{1/4}\sqrt{\Gamma_{\rho}}
\sim0.1\sqrt{\Gamma_{\rho}}\,,
\end{equation}
where 
\begin{equation}
\Gamma_{\rho}=\frac{3\:y_{N}^{2}}{64\pi}m_{\rho}\,.
\end{equation}

Using \eq{yN2} we can derive the bound
\begin{equation}
T_{RH} \lesssim 3 \times10^8\,\mathrm{GeV} \sqrt{\left(\frac{\lambda_\sigma}{10^{-7}}\right) 
\left(\frac{f_{a}}{10^{12}\:\mathrm{GeV}}\right)}
\equiv T_{RH}^{\mathrm{{max}}}\,.
\label{eq:TRH}
\end{equation}

\subsection{Non-adiabatic primordial fluctuations of axions}
\label{sec:fluctu}

Since inflation is driven by the radial part of the axion field, the PQ
symmetry is always broken during inflation and the axion acquires isothermal (more precisely isocurvature) fluctuations~\cite{Axenides:1983hj,Linde:1985yf,Linde:1991km,Linde:1984ti,Seckel:1985tj,Turner:1990uz}. In general, these are given by~\cite{Fairbairn:2014zta}
\begin{equation}
\beta_\mathrm{iso}=\left(1+\frac{\pi f_{a,\star}^{2}\overline{\theta_{i}^{2}}}{\epsilon(\rho)}
\right)^{-1}\leq0.038\,,
\label{eq:alpha}
\end{equation}
with $f_{a,\star}$ being the effective scale of PQ symmetry breaking and $\overline{\theta_{i}}$ is the spatially averaged misalignment angle.  The upper bound is the current experimental limit (95\% confidence level) at $k=0.05$ Mpc$^{-1}$~\cite{Ade:2015lrj}.  Assuming that axion DM is produced via the misalignment mechanism~\cite{Preskill:1982cy,Abbott:1982af,Dine:1982ah},  $\overline{\theta_{i}}$ enters as well in the expression of the axion relic density:
\begin{equation}
\Omega_{a}h^{2}=0.1199\:\left(\frac{\overline{\theta_{i}^{2}}}{0.28}
\right)\left(\frac{f_{a}}{10^{12}\, \mathrm{GeV}}\right)^{7/6}\,.
\label{eq:DM}
\end{equation}

\begin{figure}[t]
\centering
\includegraphics[clip,width=0.45\textwidth]{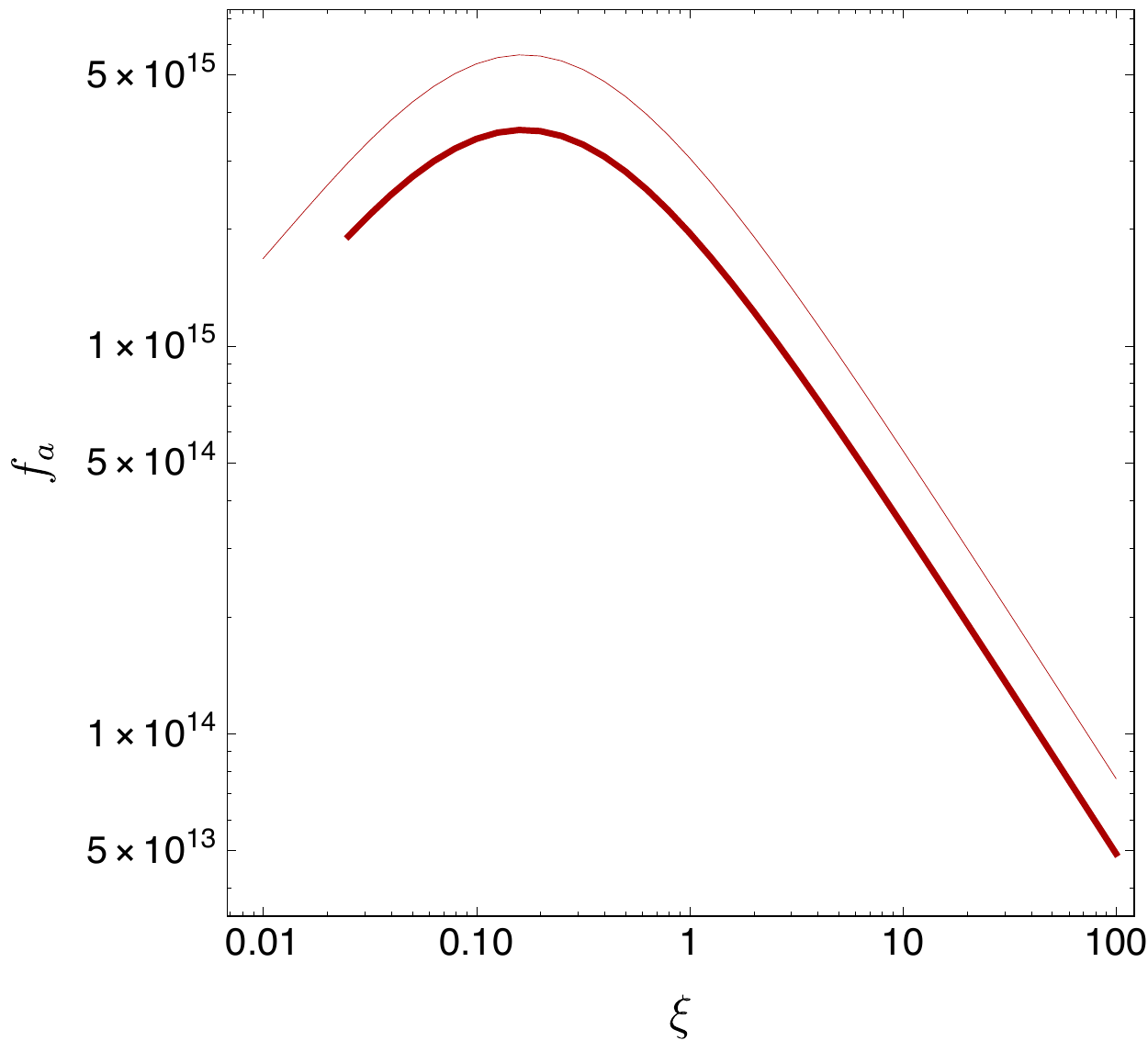}\caption{
Upper bound on the PQ scale $f_a$ as a function of the non-minimal 
coupling $\xi$ due to isocurvature and DM relic abundance constraints. The two lines are for for $N=50$ e-foldings (thick) and $N=60$ e-foldings (thin). $n_s$ is within the 68\% confidence level of {\sc Planck}'s measurement. }\label{fig:favsxi}
\end{figure}

In the standard scenario where  inflation is unrelated to axions $f_{a,\star}\equiv f_a$, and the bound \eq{alpha} favors a large PQ breaking scale. However in our case the effective scale is given by the inflaton field value during inflation, $\rho_\star$, which is trans-Planckian  for $\xi\lesssim 10^2$. Given that $\rho_\star \gg f_a$, $f_a$ does not have a direct impact on the isocurvature perturbations and enters only indirectly via \eq{DM}. Using eqs.~(\ref{eq:alpha}) and (~\ref{eq:DM}) we obtain un upper bound on $f_a$ for a given $\xi$. 
In \fig{favsxi}  we depict the predicted values of the maximal allowed value of $f_a$ as a function of the minimal coupling $\xi$ for $N=50$ and $N=60$ e-foldings. As in \fig{lvsxi}, $n_s$ is constrained to be  within the 68\% confidence level of {\sc Planck}'s measurement \cite{Ade:2015lrj}. The obtained limit on $f_a$ is compatible with the natural parameter space of axion DM.
Finally, note that for our choice of parameters, namely $f_{a}\sim10^{12}$ GeV, the PQ symmetry is
not restored at the end of  reheating since $T_{RH} \ll f_a$, see \eq{TRH}.

\section{Baryogenesis}
\label{sec:lepto}

In \eq{TRH} we found that the maximum reheat temperature obtained from inflaton's decays 
to RH neutrinos is of order  $ 10^{8}-10^9$ GeV. This allows us to implement 
baryogenesis via non-thermal leptogenesis \cite{Lazarides:1991wu}.  Assuming 
hierarchical RH neutrino masses, the lepton-to-entropy ratio \cite{Asaka:1999yd} 
is given by 
\begin{equation}
\eta_L \simeq - 10^{-5} \left(\frac{T_{RH}}{10^9 \,\mathrm{GeV}} \right)   
\left(\frac{M_N}{m_\rho} \right)\,,
\end{equation}
and the observed baryon asymmetry is related to $\eta_L$ via the usual relation 
$\eta_B  \simeq 10^{-10} \simeq -\tfrac{8}{23} \eta_L$. This leads to
\begin{equation}
M_N \simeq 0.3 \left(\frac{10^7\, \mathrm{GeV}}{T_{RH}} \right) m_\rho\,.
\end{equation}

Since $m_\rho \simeq B f_a$, we find that for $\xi\sim 1$ and $T_{RH} \sim 10^7$ 
GeV, the heavy RH neutrino with mass of the order of $10^8$ GeV can give rise to the 
observed baryon asymmetry via
non-thermal leptogenesis.

\section{GUT Monopoles}
\label{sec:monopoles}

The $SU(5)$ symmetry breaks to the SM when the effective mass squared term of 
$\Phi$, $-M_{GUT}^2+\kappa_\Phi\rho^2$, in the effective potential becomes of 
order  $-T_H^2$, with $T_H$ being the Hawking-Gibbons temperature, $T_H\equiv 
H/(2\pi)$. We want to make sure that $\Phi$ is pushed away from its origin 
during inflation. This must occur at the early stage of inflation to ensure that 
monopoles are adequately inflated away \cite{ShafiVilenkinSU5}.  In the limit 
$\kappa_\Phi \to 0$,  this translates as a lower bound on the unification scale,
\begin{equation}
 M_{GUT} > (1+\xi \rho_i^2) T_H\,.
\label{eq:mgutbound}
\end{equation} 
where $\rho_i$ is the starting value of the inflaton field. 
$\xi\rho_i^2$ varies from $\approx 0$ to $\approx 70$ in the range $\xi \in [0,100]$.
Using the result in \eq{Hinf}, we find that  the equation above yields 
$M_{GUT}\gtrsim 7 \times 10^{14}$ GeV, which is less constraining than the limit 
from proton lifetime. If $\kappa_\Phi<0$ then this condition is even easier to 
satisfy. However, in the case where  $\kappa_\Phi>0$ we will have an upper 
limit on  $\kappa_\Phi$ to ensure that 
$(M_{GUT}^2-|\kappa_\Phi|\rho_i^2)>T_H^2$ and monopoles are properly inflated 
away. For realistic $M_{GUT}$ values, we find that   $\kappa_\Phi\lesssim
10^{-7}$ for $\xi=10^{-2}$ and $\kappa_\Phi\lesssim10^{-5}$ for $\xi=10^{2}$, which is in agreement with our initial assumptions on the 
smallness of  $\kappa_\phi$.

\section{Summary and conclusions}
\label{sec:end}

We have presented a realistic grand unified theory based on $SU(5)\times U(1)_{PQ}$ which consistently addresses
multiple outstanding BSM problems. 
Fermion masses and mixings are accounted for in a renormalizable fashion, and precise gauge coupling unification
is achieved. With the unification scale predicted
to lie around $10^{16}$ GeV, proton lifetime is predicted to be in the range $8 \times 10^{34}-3 \times 10^{35}$ yr and should be accessible in the next generation detectors. The effective Higgs potential is automatically stabilized thanks to the physics used to implement gauge coupling unification.\\

The QCD axion is our candidate for the dark matter in the
universe. The axion field plays several roles in our model.
The radial component of the axion field drives inflation by exploiting
a non-minimal coupling to gravity, reheating proceeds from the axion
field coupling to RH neutrinos, and the observed baryon
asymmetry arises via non-thermal leptogenesis. The coupling with RH neutrinos inducing small
neutrino masses via the see-saw mechanism. The isocurvature fluctuations are
adequately suppressed and the axion decay constant $f_a$ lies in the
desired range of $10^{11}-10^{12}$ GeV. The model predicts a tensor
to scalar $r$ in an observable range, $r= \mathrm{few} \times 10^{-3}$. We finally comment that the discussion can be extended to realistic $SO(10)$ models with a suitable intermediate scale such as $SU(4)\times SU(2)\times SU(2)$ or   $SU(3)\times SU(2)\times SU(2) \times U(1)$.

\section*{Acknowledgments}
S.B.~thanks the ``Roland Gustafssons Stiftelse f{\"o}r teoretisk fysik'' for financial support. Q.S is supported in part by a DOE grant No DE-SC 0013880.

\bibliographystyle{utphys}
\bibliography{refs.bib}

\providecommand{\href}[2]{#2}\begingroup\raggedright\begin{thebibliography}{10}

\bibitem{ShafiVilenkinSU5}
Q.~Shafi and A.~Vilenkin, ``{Inflation with SU(5)},''
\href{http://dx.doi.org/10.1103/PhysRevLett.52.691}{{\em Phys. Rev. Lett.}
  {\bfseries 52} (1984) 691--694}.

\bibitem{Rehman:2008qs}
M.~U. Rehman, Q.~Shafi, and J.~R. Wickman, ``{GUT Inflation and Proton Decay
  after WMAP5},'' \href{http://dx.doi.org/10.1103/PhysRevD.78.123516}{{\em
  Phys. Rev.} {\bfseries D78} (2008) 123516},
\href{http://arxiv.org/abs/0810.3625}{{\ttfamily arXiv:0810.3625 [hep-ph]}}.

\bibitem{Wilczek:1977pj}
F.~Wilczek, ``{Problem of Strong P and T Invariance in the Presence of
  Instantons},''
\href{http://dx.doi.org/10.1103/PhysRevLett.40.279}{{\em Phys. Rev. Lett.}
  {\bfseries 40} (1978) 279--282}.

\bibitem{Weinberg:1977ma}
S.~Weinberg, ``{A New Light Boson?},''
\href{http://dx.doi.org/10.1103/PhysRevLett.40.223}{{\em Phys. Rev. Lett.}
  {\bfseries 40} (1978) 223--226}.

\bibitem{Pi:1984pv}
S.-Y. Pi, ``{Inflation Without Tears},''
\href{http://dx.doi.org/10.1103/PhysRevLett.52.1725}{{\em Phys. Rev. Lett.}
  {\bfseries 52} (1984) 1725--1728}.

\bibitem{Okada:2014lxa}
N.~Okada, V.~N. Şenoğuz, and Q.~Shafi, ``{The Observational Status of Simple
  Inflationary Models: an Update},''
  \href{http://dx.doi.org/10.3906/fiz-1505-7}{{\em Turk. J. Phys.} {\bfseries
  40} no.~2, (2016) 150--162},
\href{http://arxiv.org/abs/1403.6403}{{\ttfamily arXiv:1403.6403 [hep-ph]}}.

\bibitem{Langacker:1986rj}
P.~Langacker, R.~D. Peccei, and T.~Yanagida, ``{Invisible Axions and Light
  Neutrinos: Are They Connected?},''
\href{http://dx.doi.org/10.1142/S0217732386000683}{{\em Mod. Phys. Lett.}
  {\bfseries A1} (1986) 541}.

\bibitem{Shin:1987xc}
M.~Shin, ``{Light Neutrino Masses and Strong {CP} Problem},''
  \href{http://dx.doi.org/10.1103/PhysRevLett.59.2515}{{\em Phys. Rev. Lett.}
  {\bfseries 59} (1987) 2515}.
[Erratum: Phys. Rev. Lett.60,383(1988)].

\bibitem{Davoudiasl:2004be}
H.~Davoudiasl, R.~Kitano, T.~Li, and H.~Murayama, ``{The New minimal standard
  model},'' \href{http://dx.doi.org/10.1016/j.physletb.2005.01.026}{{\em Phys.
  Lett.} {\bfseries B609} (2005) 117--123},
\href{http://arxiv.org/abs/hep-ph/0405097}{{\ttfamily arXiv:hep-ph/0405097
  [hep-ph]}}.

\bibitem{Shaposhnikov:2006xi}
M.~Shaposhnikov and I.~Tkachev, ``{The nuMSM, inflation, and dark matter},''
  \href{http://dx.doi.org/10.1016/j.physletb.2006.06.063}{{\em Phys. Lett.}
  {\bfseries B639} (2006) 414--417},
\href{http://arxiv.org/abs/hep-ph/0604236}{{\ttfamily arXiv:hep-ph/0604236
  [hep-ph]}}.

\bibitem{Boucenna:2014uma}
S.~M. Boucenna, S.~Morisi, Q.~Shafi, and J.~W.~F. Valle, ``{Inflation and
  majoron dark matter in the seesaw mechanism},''
  \href{http://dx.doi.org/10.1103/PhysRevD.90.055023}{{\em Phys. Rev.}
  {\bfseries D90} no.~5, (2014) 055023},
\href{http://arxiv.org/abs/1404.3198}{{\ttfamily arXiv:1404.3198 [hep-ph]}}.

\bibitem{Bertolini:2014aia}
S.~Bertolini, L.~Di~Luzio, H.~Kolešová, and M.~Malinský, ``{Massive
  neutrinos and invisible axion minimally connected},''
  \href{http://dx.doi.org/10.1103/PhysRevD.91.055014}{{\em Phys. Rev.}
  {\bfseries D91} no.~5, (2015) 055014},
\href{http://arxiv.org/abs/1412.7105}{{\ttfamily arXiv:1412.7105 [hep-ph]}}.

\bibitem{Clarke:2015bea}
J.~D. Clarke and R.~R. Volkas, ``{Technically natural nonsupersymmetric model
  of neutrino masses, baryogenesis, the strong CP problem, and dark matter},''
  \href{http://dx.doi.org/10.1103/PhysRevD.93.035001}{{\em Phys. Rev.}
  {\bfseries D93} no.~3, (2016) 035001},
  \href{http://arxiv.org/abs/1509.07243}{{\ttfamily arXiv:1509.07243
  [hep-ph]}}.
[Phys. Rev.D93,035001(2016)].

\bibitem{Ahn:2015pia}
Y.~H. Ahn and E.~J. Chun, ``{Minimal Models for Axion and Neutrino},''
  \href{http://dx.doi.org/10.1016/j.physletb.2015.11.067}{{\em Phys. Lett.}
  {\bfseries B752} (2016) 333--337},
\href{http://arxiv.org/abs/1510.01015}{{\ttfamily arXiv:1510.01015 [hep-ph]}}.

\bibitem{Salvio:2015cja}
A.~Salvio, ``{A Simple Motivated Completion of the Standard Model below the
  Planck Scale: Axions and Right-Handed Neutrinos},''
  \href{http://dx.doi.org/10.1016/j.physletb.2015.03.015}{{\em Phys. Lett.}
  {\bfseries B743} (2015) 428--434},
\href{http://arxiv.org/abs/1501.03781}{{\ttfamily arXiv:1501.03781 [hep-ph]}}.

\bibitem{Ballesteros:2016euj}
G.~Ballesteros, J.~Redondo, A.~Ringwald, and C.~Tamarit, ``{Unifying inflation
  with the axion, dark matter, baryogenesis and the seesaw mechanism},''
  \href{http://dx.doi.org/10.1103/PhysRevLett.118.071802}{{\em Phys. Rev.
  Lett.} {\bfseries 118} no.~7, (2017) 071802},
\href{http://arxiv.org/abs/1608.05414}{{\ttfamily arXiv:1608.05414 [hep-ph]}}.

\bibitem{Georgi:1974sy}
H.~Georgi and S.~L. Glashow, ``{Unity of All Elementary Particle Forces},''
\href{http://dx.doi.org/10.1103/PhysRevLett.32.438}{{\em Phys. Rev. Lett.}
  {\bfseries 32} (1974) 438--441}.

\bibitem{Peccei:1977ur}
R.~D. Peccei and H.~R. Quinn, ``{Constraints Imposed by CP Conservation in the
  Presence of Instantons},''
\href{http://dx.doi.org/10.1103/PhysRevD.16.1791}{{\em Phys. Rev.} {\bfseries
  D16} (1977) 1791--1797}.

\bibitem{Zhitnitsky:1980tq}
A.~R. Zhitnitsky, ``{On Possible Suppression of the Axion Hadron Interactions.
  (In Russian)},'' {\em Sov. J. Nucl. Phys.} {\bfseries 31} (1980) 260.
[Yad. Fiz.31,497(1980)].

\bibitem{Dine:1981rt}
M.~Dine, W.~Fischler, and M.~Srednicki, ``{A Simple Solution to the Strong CP
  Problem with a Harmless Axion},''
\href{http://dx.doi.org/10.1016/0370-2693(81)90590-6}{{\em Phys. Lett.}
  {\bfseries 104B} (1981) 199--202}.

\bibitem{Wise:1981ry}
M.~B. Wise, H.~Georgi, and S.~L. Glashow, ``{SU(5) and the Invisible Axion},''
\href{http://dx.doi.org/10.1103/PhysRevLett.47.402}{{\em Phys. Rev. Lett.}
  {\bfseries 47} (1981) 402}.

\bibitem{Chikashige:1980ui}
Y.~Chikashige, R.~N. Mohapatra, and R.~D. Peccei, ``{Are There Real Goldstone
  Bosons Associated with Broken Lepton Number?},''
\href{http://dx.doi.org/10.1016/0370-2693(81)90011-3}{{\em Phys. Lett.}
  {\bfseries 98B} (1981) 265--268}.

\bibitem{Gelmini:1980re}
G.~B. Gelmini and M.~Roncadelli, ``{Left-Handed Neutrino Mass Scale and
  Spontaneously Broken Lepton Number},''
\href{http://dx.doi.org/10.1016/0370-2693(81)90559-1}{{\em Phys. Lett.}
  {\bfseries 99B} (1981) 411--415}.

\bibitem{Georgi:1979df}
H.~Georgi and C.~Jarlskog, ``{A New Lepton - Quark Mass Relation in a Unified
  Theory},''
\href{http://dx.doi.org/10.1016/0370-2693(79)90842-6}{{\em Phys. Lett.}
  {\bfseries 86B} (1979) 297--300}.

\bibitem{Kalyniak:1982pt}
P.~Kalyniak and J.~N. Ng, ``{Symmetry Breaking Patterns in SU(5) With
  Nonminimal Higgs Fields},''
\href{http://dx.doi.org/10.1103/PhysRevD.26.890}{{\em Phys. Rev.} {\bfseries
  D26} (1982) 890}.

\bibitem{Eckert:1983bn}
P.~Eckert, J.~M. Gerard, H.~Ruegg, and T.~Schucker, ``{Minimization of the
  SU(5) Invariant Scalar Potential for the Fortyfive-dimensional
  Representation},''
\href{http://dx.doi.org/10.1016/0370-2693(83)91308-4}{{\em Phys. Lett.}
  {\bfseries 125B} (1983) 385--388}.

\bibitem{Dorsner:2006dj}
I.~Dorsner and P.~Fileviez~Perez, ``{Unification versus proton decay in
  SU(5)},'' \href{http://dx.doi.org/10.1016/j.physletb.2006.09.034}{{\em Phys.
  Lett.} {\bfseries B642} (2006) 248--252},
\href{http://arxiv.org/abs/hep-ph/0606062}{{\ttfamily arXiv:hep-ph/0606062
  [hep-ph]}}.

\bibitem{PDG2016}
{\bfseries Particle Data Group} Collaboration, C.~Patrignani {\em et~al.},
  ``{Review of Particle Physics},''
\href{http://dx.doi.org/10.1088/1674-1137/40/10/100001}{{\em Chin. Phys.}
  {\bfseries C40} no.~10, (2016) 100001}.

\bibitem{Abe:2011ts}
K.~Abe {\em et~al.}, ``{Letter of Intent: The Hyper-Kamiokande Experiment ---
  Detector Design and Physics Potential ---},''
\href{http://arxiv.org/abs/1109.3262}{{\ttfamily arXiv:1109.3262 [hep-ex]}}.

\bibitem{Buttazzo:2013uya}
D.~Buttazzo, G.~Degrassi, P.~P. Giardino, G.~F. Giudice, F.~Sala, A.~Salvio,
  and A.~Strumia, ``{Investigating the near-criticality of the Higgs boson},''
  \href{http://dx.doi.org/10.1007/JHEP12(2013)089}{{\em JHEP} {\bfseries 12}
  (2013) 089},
\href{http://arxiv.org/abs/1307.3536}{{\ttfamily arXiv:1307.3536 [hep-ph]}}.

\bibitem{Machacek:1983tz}
M.~E. Machacek and M.~T. Vaughn, ``{Two Loop Renormalization Group Equations in
  a General Quantum Field Theory. 1. Wave Function Renormalization},''
\href{http://dx.doi.org/10.1016/0550-3213(83)90610-7}{{\em Nucl. Phys.}
  {\bfseries B222} (1983) 83--103}.

\bibitem{Machacek:1984zw}
M.~E. Machacek and M.~T. Vaughn, ``{Two Loop Renormalization Group Equations in
  a General Quantum Field Theory. 3. Scalar Quartic Couplings},''
\href{http://dx.doi.org/10.1016/0550-3213(85)90040-9}{{\em Nucl. Phys.}
  {\bfseries B249} (1985) 70--92}.

\bibitem{Ford:1992pn}
C.~Ford, I.~Jack, and D.~R.~T. Jones, ``{The Standard model effective potential
  at two loops},'' \href{http://dx.doi.org/10.1016/0550-3213(92)90165-8,
  10.1016/S0550-3213(97)00532-4}{{\em Nucl. Phys.} {\bfseries B387} (1992)
  373--390}, \href{http://arxiv.org/abs/hep-ph/0111190}{{\ttfamily
  arXiv:hep-ph/0111190 [hep-ph]}}.
[Erratum: Nucl. Phys.B504,551(1997)].

\bibitem{Arason:1991ic}
H.~Arason, D.~J. Castano, B.~Keszthelyi, S.~Mikaelian, E.~J. Piard, P.~Ramond,
  and B.~D. Wright, ``{Renormalization group study of the standard model and
  its extensions. 1. The Standard model},''
\href{http://dx.doi.org/10.1103/PhysRevD.46.3945}{{\em Phys. Rev.} {\bfseries
  D46} (1992) 3945--3965}.

\bibitem{Barger:1992ac}
V.~D. Barger, M.~S. Berger, and P.~Ohmann, ``{Supersymmetric grand unified
  theories: Two loop evolution of gauge and Yukawa couplings},''
  \href{http://dx.doi.org/10.1103/PhysRevD.47.1093}{{\em Phys. Rev.} {\bfseries
  D47} (1993) 1093--1113},
\href{http://arxiv.org/abs/hep-ph/9209232}{{\ttfamily arXiv:hep-ph/9209232
  [hep-ph]}}.

\bibitem{Luo:2002ey}
M.-x. Luo and Y.~Xiao, ``{Two loop renormalization group equations in the
  standard model},''
  \href{http://dx.doi.org/10.1103/PhysRevLett.90.011601}{{\em Phys. Rev. Lett.}
  {\bfseries 90} (2003) 011601},
\href{http://arxiv.org/abs/hep-ph/0207271}{{\ttfamily arXiv:hep-ph/0207271
  [hep-ph]}}.

\bibitem{ATLAS:2014wva}
{\bfseries ATLAS, CDF, CMS, D0} Collaboration, ``{First combination of Tevatron
  and LHC measurements of the top-quark mass},''
\href{http://arxiv.org/abs/1403.4427}{{\ttfamily arXiv:1403.4427 [hep-ex]}}.

\bibitem{Aad:2015zhl}
{\bfseries ATLAS, CMS} Collaboration, G.~Aad {\em et~al.}, ``{Combined
  Measurement of the Higgs Boson Mass in $pp$ Collisions at $\sqrt{s}=7$ and 8
  TeV with the ATLAS and CMS Experiments},''
  \href{http://dx.doi.org/10.1103/PhysRevLett.114.191803}{{\em Phys. Rev.
  Lett.} {\bfseries 114} (2015) 191803},
\href{http://arxiv.org/abs/1503.07589}{{\ttfamily arXiv:1503.07589 [hep-ex]}}.

\bibitem{Okada:2013vxa}
N.~Okada and Q.~Shafi, ``{Observable Gravity Waves From $U(1)_{B-L}$ Higgs and
  Coleman-Weinberg Inflation},''
\href{http://arxiv.org/abs/1311.0921}{{\ttfamily arXiv:1311.0921 [hep-ph]}}.

\bibitem{Liddle:1992wi}
A.~R. Liddle and D.~H. Lyth, ``{COBE, gravitational waves, inflation and
  extended inflation},''
  \href{http://dx.doi.org/10.1016/0370-2693(92)91393-N}{{\em Phys. Lett.}
  {\bfseries B291} (1992) 391--398},
\href{http://arxiv.org/abs/astro-ph/9208007}{{\ttfamily arXiv:astro-ph/9208007
  [astro-ph]}}.

\bibitem{Liddle:1993fq}
A.~R. Liddle and D.~H. Lyth, ``{The Cold dark matter density perturbation},''
  \href{http://dx.doi.org/10.1016/0370-1573(93)90114-S}{{\em Phys. Rept.}
  {\bfseries 231} (1993) 1--105},
\href{http://arxiv.org/abs/astro-ph/9303019}{{\ttfamily arXiv:astro-ph/9303019
  [astro-ph]}}.

\bibitem{Ade:2015lrj}
{\bfseries Planck} Collaboration, P.~A.~R. Ade {\em et~al.}, ``{Planck 2015
  results. XX. Constraints on inflation},''
  \href{http://dx.doi.org/10.1051/0004-6361/201525898}{{\em Astron. Astrophys.}
  {\bfseries 594} (2016) A20},
\href{http://arxiv.org/abs/1502.02114}{{\ttfamily arXiv:1502.02114
  [astro-ph.CO]}}.

\bibitem{Axenides:1983hj}
M.~Axenides, R.~H. Brandenberger, and M.~S. Turner, ``{Development of Axion
  Perturbations in an Axion Dominated Universe},''
\href{http://dx.doi.org/10.1016/0370-2693(83)90586-5}{{\em Phys. Lett.}
  {\bfseries 126B} (1983) 178--182}.

\bibitem{Linde:1985yf}
A.~D. Linde, ``{Generation of Isothermal Density Perturbations in the
  Inflationary Universe},''
\href{http://dx.doi.org/10.1016/0370-2693(85)90436-8}{{\em Phys. Lett.}
  {\bfseries 158B} (1985) 375--380}.

\bibitem{Linde:1991km}
A.~D. Linde, ``{Axions in inflationary cosmology},''
\href{http://dx.doi.org/10.1016/0370-2693(91)90130-I}{{\em Phys. Lett.}
  {\bfseries B259} (1991) 38--47}.

\bibitem{Linde:1984ti}
A.~D. Linde, ``{Generation of isothermal density perturbations in the
  inflationary universe},'' {\em JETP Lett.} {\bfseries 40} (1984) 1333--1336.
[Pisma Zh. Eksp. Teor. Fiz.40,496(1984)].

\bibitem{Seckel:1985tj}
D.~Seckel and M.~S. Turner, ``{Isothermal Density Perturbations in an Axion
  Dominated Inflationary Universe},''
\href{http://dx.doi.org/10.1103/PhysRevD.32.3178}{{\em Phys. Rev.} {\bfseries
  D32} (1985) 3178}.

\bibitem{Turner:1990uz}
M.~S. Turner and F.~Wilczek, ``{Inflationary axion cosmology},''
\href{http://dx.doi.org/10.1103/PhysRevLett.66.5}{{\em Phys. Rev. Lett.}
  {\bfseries 66} (1991) 5--8}.

\bibitem{Fairbairn:2014zta}
M.~Fairbairn, R.~Hogan, and D.~J.~E. Marsh, ``{Unifying inflation and dark
  matter with the Peccei-Quinn field: observable axions and observable
  tensors},'' \href{http://dx.doi.org/10.1103/PhysRevD.91.023509}{{\em Phys.
  Rev.} {\bfseries D91} no.~2, (2015) 023509},
\href{http://arxiv.org/abs/1410.1752}{{\ttfamily arXiv:1410.1752 [hep-ph]}}.

\bibitem{Preskill:1982cy}
J.~Preskill, M.~B. Wise, and F.~Wilczek, ``{Cosmology of the Invisible
  Axion},''
\href{http://dx.doi.org/10.1016/0370-2693(83)90637-8}{{\em Phys. Lett.}
  {\bfseries 120B} (1983) 127--132}.

\bibitem{Abbott:1982af}
L.~F. Abbott and P.~Sikivie, ``{A Cosmological Bound on the Invisible Axion},''
\href{http://dx.doi.org/10.1016/0370-2693(83)90638-X}{{\em Phys. Lett.}
  {\bfseries 120B} (1983) 133--136}.

\bibitem{Dine:1982ah}
M.~Dine and W.~Fischler, ``{The Not So Harmless Axion},''
\href{http://dx.doi.org/10.1016/0370-2693(83)90639-1}{{\em Phys. Lett.}
  {\bfseries 120B} (1983) 137--141}.

\bibitem{Lazarides:1991wu}
G.~Lazarides and Q.~Shafi, ``{Origin of matter in the inflationary
  cosmology},''
\href{http://dx.doi.org/10.1016/0370-2693(91)91090-I}{{\em Phys. Lett.}
  {\bfseries B258} (1991) 305--309}.

\bibitem{Asaka:1999yd}
T.~Asaka, K.~Hamaguchi, M.~Kawasaki, and T.~Yanagida, ``{Leptogenesis in
  inflaton decay},''
  \href{http://dx.doi.org/10.1016/S0370-2693(99)01020-5}{{\em Phys. Lett.}
  {\bfseries B464} (1999) 12--18},
\href{http://arxiv.org/abs/hep-ph/9906366}{{\ttfamily arXiv:hep-ph/9906366
  [hep-ph]}}.

\end{thebibliography}\endgroup

\end{document}